# Electronic Transport in Composites of Graphite Oxide with Carbon Nanotubes

Viera Skákalová[1,2*], Viliam Vretenár[1,3], Ľubomír Kopera[5], Peter Kotrusz[1], Clemens Mangler[2], Marcel Meško[1], Jannik C. Meyer[2] and Martin Hulman[1,4]

[1]*Danubia NanoTech s.r.o., Ilkovičova 3, 84104 Bratislava, Slovakia*

[2]*University of Vienna, Faculty of Physics, Group Physics of nanostructured materials, Boltzmanngasse 5, 1090 Vienna, Austria*

[3]*Institute of Physics, Slovak Academy of Science, Dúbravská cesta 9, 845 11 Bratislava, Slovakia*

[4]*International Laser Center, Ilkovičova 3, 84104 Bratislava, Slovakia*

[5] *Institute of Electrical Engineering, Slovak Academy of Science, Dúbravská cesta 9, 845 11 Bratislava, Slovakia*

**Abstract**

We show that the presence of electrically insulating graphite oxide (GO) within a single wall carbon nanotube (SWCNT) network strongly enhances electrical conductivity, whereas reduced graphite oxide, even though electrically conductive, suppresses electrical conductivity within a composite network with SWCNTs. Measurements of Young´s modulus and of Raman spectra strongly support our interpretation of the "indirect" role of the oxide groups, present in GO within the SWCNT-GO composite, through electronic doping of metallic SWCNTs.

*Tel.: +43 664 60277 51328
*E-mail: viera.skakalova@univie.ac.at

1. **Introduction**

Electronic transport in low-dimensional graphitic carbon structures is determined firstly by the dimensionality (0, 1 or 2) of the system. The significance of electronic transport for the zero-dimensional carbon form – fullerene - is negligible, as these tend to avoid double bonds in the pentagonal rings, resulting in poor electron delocalisation, whereas carbon nanotubes, with dimensionality "quasi-one", as well as entirely 2-dimensional graphene, are well appreciated for their electrical conduction with ballistic transport. Regarding carbon nanotubes, it is network of carbon nanotubes that is most likely to be used in technological applications, as electronic circuits are rarely formed from individual nanotubes. But then we have to ask: to what extent transport properties entirely bound to an individual carbon nanotube can still be preserved in a macroscopic network? Various factors control electronic transport through a nanotube network, such as the network´s packing density, the length and orientation of the tubes and, most importantly, the interactions between the nanotubes at their contacts [1,2]. For current to efficiently pass along the network, strong inter-tube interactions are necessary. Electronic doping is a powerful way to enhance the tunnelling of charges through the inter-tube contacts [3,4].

In that sense, the same situation holds for graphene as well; also here the phenomena appreciated in graphene´s electronic and magneto-transport properties [5,6] are confined inside a single graphene crystal, while they are hardly expected to be observed in an ensemble of weakly bound graphene flakes. Graphene-flake network is relevant for many application of graphene in the large-scale. The "scotch-tape" method, so successful in the first experiments, cannot be used anymore; instead, techniques for large-scale preparation of graphene had to be developed. One of them uses exfoliation of graphitic powders through chemical oxidation [7,8], that produces graphitic sheets with high efficiency; but the oxide groups covalently attached to the graphene´s surface (graphite oxide (GO)) cause strong electron localization, shutting down the charge transport. Electrical conductivity recovers, after the oxide groups are reduced. Measuring electronic transport through devices fabricated of a single layer of rGO Kaiser et al. [9,10] showed that the 2-dimansional variable-range hopping (VRH) is employed at electronic transport, whereas 3-dimensional VRH better fitted when a thin GO film based device was investigated [11,12].

The main aim of this work is to investigate the electronic transport mechanism in composite networks of 1-dimensional SWCNTs mixed with 2-dimensional flakes of graphite oxide (in its oxidized and reduced form), through measuring the temperature dependence of electrical conductivity.

As a surprising result, we find that a composite network of SWCNTs and GO (which is electrically insulating) shows a higher conductivity than pure SWCNTs or GO by itself. This is interpreted as an effect of electron transfer (doping) from GO to SWCNTs, as evidenced also by Raman spectroscopy.

2. Experimental

We have studied electronic properties of networks of SWCNTs, rGO, and, in particular, of composite networks formed by SWCNTs+GO and SWCNTs+rGO. All samples were prepared by vacuum filtration from a solution, or mixtures of solutions. Besides the electronic transport measurements at variable temperature which form the central aspect of this work, the samples were characterized by Raman spectroscopy, scanning electron microscopy, high-resolution transmission electron microscopy, and mechanical testing as detailed below.

## 2.1 Sample preparation

### 2.1.1 GO synthesis

Natural graphite powder of microcrystal grade, purity of 99.9995% and size of 2-15 μm (APS) was purchased from Alfa Aesar. Sulphuric acid (350ml) was mixed with graphite (2g) at 0-5°C for 15 minutes. Potassium permanganate (8g) and sodium nitrate (1g) were added portion-wise at 0°C and stirred for 30 minutes, then for 30 minutes at 35°C. Water (250ml) was added via dropping funnel and the reaction mixture was heated to 98°C for 3 hours. The reaction was terminated by adding 500 ml of water deionized and 40ml of 30% $H_2O_2$. The mixture was filtered off through nylon filter, washed with diluted HCl (10%) in order to remove metal ions and then with water until pH of filtrate is about 7, and dried at 75°C.

### 2.1.2 Chemical reduction

Deionized water (150ml) was added to GO (1g) and vigorously stirred for 24 hours at room temperature. The suspension was then sonicated in ultrasonic bath cleaner for 3 hours, then sonicated with fingers for 30 minutes and, finally, for 1 hour in bath sonicator. The mixture was mixed with ammonia (1.5ml) and hydrazine monohydrate (3ml) and stirred vigorously at 85°C for 24 hours under reflux condenser. After cooling, the suspension was filtered off through nylon filter, washed with deionized water (500ml) and with methanol (50ml). The cake was dried at 75°C in oven for 24 hours.

### 2.1.3 Thermal reduction

GO free-standing film was inserted into a quartz tube centred in a cylindrical furnace. After evacuation, the sample was slowly heated up to 700 °C under $Ar/H_2$ atmosphere for 2 hours.

### 2.1.4 Preparation of networks

Free-standing SWCNT and GO based papers were prepared from a suspension by vacuum filtration. For SWCNTs (prepared by the Laser Ablation method), a *N*-Methyl-2-pyrolidone (NMP) based suspension was used, whereas GO was suspended in water. The composite of SWCNT with GO was prepared by mixing 50wt% of SWCNT and 50wt% of GO in 1wt% SDS/water solution, while the composite of SWCNT with chemically reduced graphite oxide (rGO) was prepared by mixing SWCNT and rGO at same weight ratio in the NMP based suspension. The surface morphology of the papers imaged by SEM is illustrated in Fig. 1b.

### 2.2 Characterization

#### 2.2.1 Scanning electron microscopy

The samples` morphology was examined by SEM JEOL JSM-7500F. Figure 1 presents four samples studied in this work: (a) SWCNTs showing the typical spaghetti-like morphology at approximately 10-times higher magnification than the other images containing GO flakes; (b) thermally-reduced rGO; (c) composite SWCNT-GO; and (d) composite SWCNT-rGO all of them in a similar scale. In the composites, the carbon nanotubes are well intercalated into the GO flakes.

Figure 1 clearly demonstrates an important difference in the morphologies between SWCNT-GO and SWCNT-rGO: While the SWCNTs are tightly integrated in the layers of densely packed GO sheets (Fig. 1(c)), the structure of SWCNT-rGO in Fig. 1(d) is loose. The interactions between the particles, well demonstrated in the SEM images, determine all the properties studied in this paper.

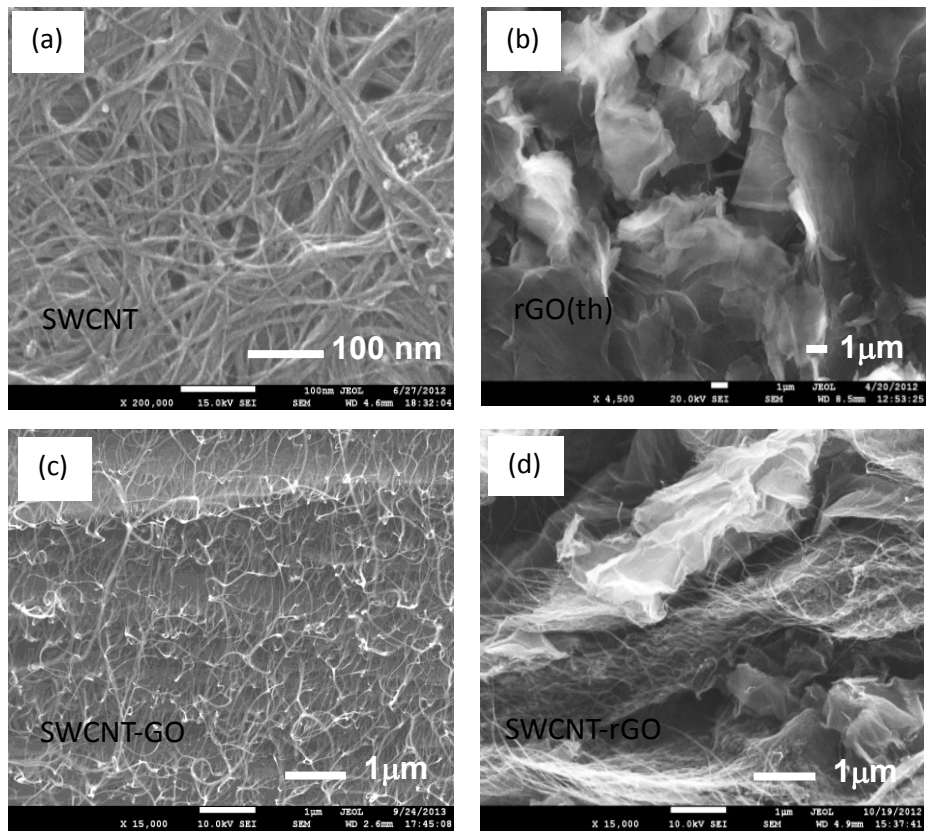

**Figure 1:** SEM images of a SWCNT paper (a), a thermally reduced -rGO (b), a composite SWCNT-GO (c) and a composite SWCNT-rGO (d).

### 2.2.2 Transmission electron microscopy

TEM investigations were carried out using a Philips CM200 microscope operating at an accelerating voltage 80 keV and selected-area diffraction (SAD) pattern taken from areas of 1 µm diameter were recorded on a CCD camera.

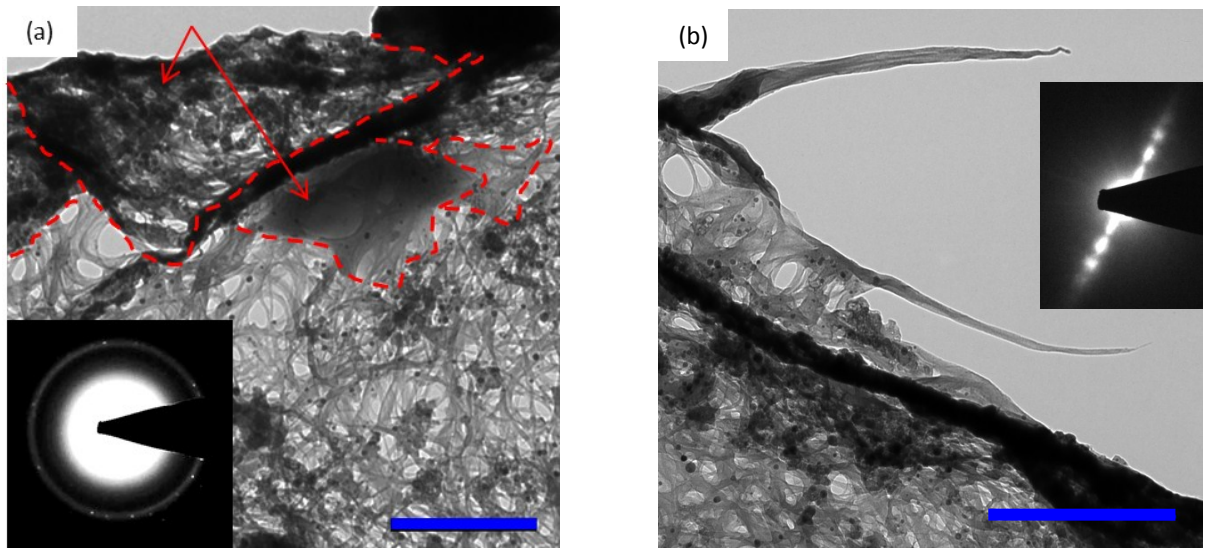

**Figure 2:** TEM images of the composite SWCNT-GO: (a) Top view of the composite; the inset shows electron diffraction, (b) Side view of a bundle of SWCNTs; the inset shows electron diffraction. Dense GO areas and large GO flakes are marked by dashed red line. The scale-bars correspond to 500 nm.

While SEM provided a picture of the cross-section of a bulk sample, for TEM a small piece of material was pulled out of the bulk and the edge layer was imaged. Imaging more detailed morphology in the transmission mode helps to better understand particle interactions within the composite. Figure 2(a) shows a top view of the composite SWCNT-GO. Here a percolating SWCNT network is mostly intercalated with GO-flakes often smaller than a micrometer size, while high density GO areas and large GO flakes are marked by dashed red line. SEM in Figure 1 indicated also significantly larger graphitic platelets. The inset shows electron diffraction; the smeared ring is formed due to a background of randomly oriented GO-flakes of a small size, whereas the bright spots decorating the ring originate from the flakes directly facing the electron beam with the distinct diffraction pattern of the hexagonal lattice of graphene. In (b), bundles of SWCNTs pulled out of the network are imaged and nicely demonstrate the typical diffraction pattern of nanotubes shown in the inset of Figure 2(b).

### 2.2.3 Raman spectroscopy

Raman spectra of SWCNT and of composite SWCNT-GO and SWCNT-rGO(ch) papers excited with a wavelength of 514 nm were obtained via the PRINCETON RESEARCH spectrometer. The spectra at 633nm excitation line were acquired using a Labram spectrometer. In both cases, microscopes were used for collecting backscattered light. The power of incident laser light was kept low enough to avoid excessive heating and thermal reduction of GO samples.

### 2.2.4 Young´s modulus measurement

The elastic Young´s modulus was determined from stress versus strain $\sigma(\varepsilon)$ characteristic as $Y = \sigma/\varepsilon$ using a home-made setup. The samples of the size 10x3 mm$^2$ were fixed between two pairs of clams connected to a micrometric screw which controlled the strain with a precision of 0.5 μm. The corresponding stress was measured by a forcemeter Scout 55, HBM (Hottinger Baldwin Messtechnik).

### 2.2.5 Electronic transport measurements

The $R(T)$ characteristics were measured by the four-probe method, within a temperature range of 25 to 310 K. Four parallel gold stripes were evaporated through a mask on samples of size 4x2 mm$^2$, with a distance of 0.5 mm. Then the samples were fixed on a Si substrate and attached to the copper block, isolated with a 50 μm thick insulating film and wire-bonded to the electrical leads of the sample holder. The technical solution of the measurement system and the sample holder`s design are described elsewhere [13]. Voltage readings from the potential leads, Pt100 thermometer and precise current shunt were controlled by a 24 bit data acquisition (DAQ) system, USB2AD from Arepoc s.r.o., Slovakia. The measurement was performed at a constant current 200 μA under ~ 50 Pa helium atmosphere. The rate of heating was 50 mK/s.

## 3. Results

### 3.1 Raman spectroscopy

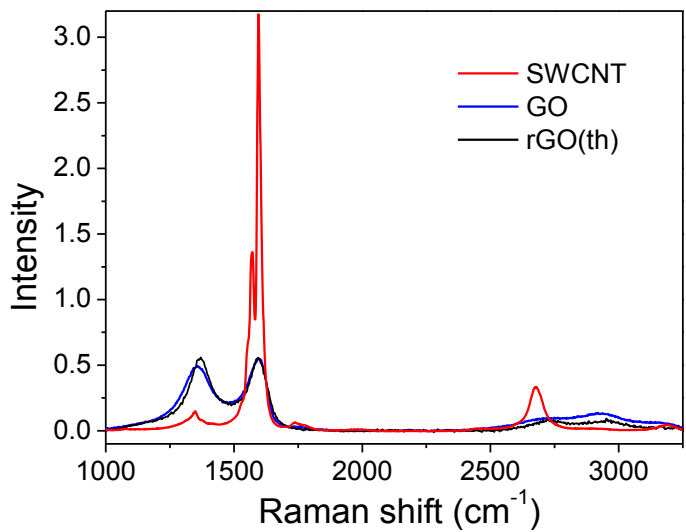

**Figure 3:** Comparison of Raman spectra of pristine SWCNT, GO and rGO(th) at excitation wavelength 514 nm.

Figure 3 compares Raman spectra of SWCNT (red line), GO (blue line) and rGO(th) (black line) at the excitation wavelength 514 nm. The resonant nature of the Raman response from the SWCNT, due to the 1D singularities in the electronic density of states, makes it strongly dominating over the Raman signal from GO and rGO. Therefore, the Raman spectra of composite samples reflect mainly the electronic structure of SWCNTs in either GO or rGO environment.

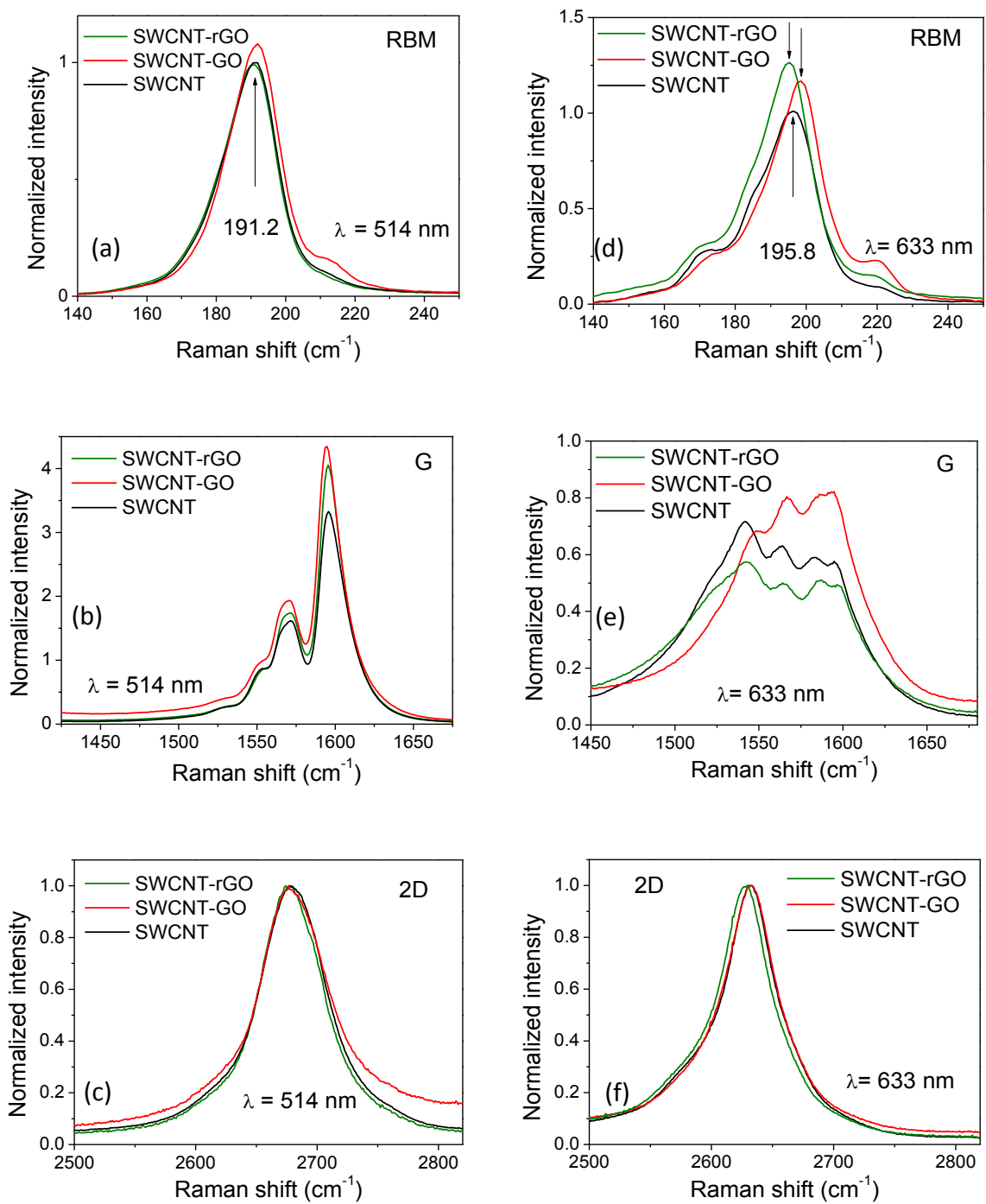

**Figure 4:** Comparison of Raman spectra of pristine SWCNT, composite SWCNT-GO and composite SWCNT-rGO at excitation wavelength 514 nm (a) RBM-modes, (c) G-modes (e) 2D-modes and, at excitation wavelength 633 nm, (b) RBM-modes, (d) G-modes (f) 2D-modes. The numbers in panels (a) and (d) denote the peak positions used for determining diameters of the tubes in resonance.

Figure 4 presents Raman spectra of the pristine SWCNTs (black lines), of the composite

SWCNT-GO (red lines) and SWCNT-rGO (green lines) at two excitation wavelengths, 514 nm and 633 nm. The main three modes (RBM, G and 2D) of SWCNT spectrum at both excitation lines ((a), (b) and (c) at 514 nm and (d), (e) and (f) at 633 nm) are plotted in separate panels: the low frequency radial breathing mode RBM in Figure 4 (a), (d), the medium frequency G-mode (in-plane C-C stretching) in Figure 4 (b), (e) and the second-order double-resonance 2D-mode in Figure 4 (c), (f).

For easier comparison, all the spectra are normalized to the 2D bands so that their intensity equals to 1. This mode was chosen for normalizing because, except of the shift in $\omega_{2D}$ due to different excitation wavelengths (at $\lambda = 514$ nm: $\omega_{2D} = 2678$ cm$^{-1}$, whereas at $\lambda = 633$ nm: $\omega_{2D} = 2632$ cm$^{-1}$), the positions and shapes of the 2D peaks for all the samples at the same excitation wavelengths stay mostly unchanged.

### 3.2 Mechanical properties

We measured the elastic modulus of SWCNT-GO and SWCNT-rGO composites, as well as in pristine SWCNT paper, as a reference. The results in Table I demonstrate that the SWCNT-GO composite is densely packed and exhibits a 50% higher Young´s modulus value than that in pristine SWCNTs. On the other hand, the SWCNT-rGO composite has lower volume density accompanied by very loose elastic properties. Unfortunately, we did not manage to measure the value of the elastic modulus of the rGO(th) paper because it became too brittle to resist the mechanical strain necessary for determining the value of Young´s modulus.

**Table I** Volume density and Young´s modulus for various samples.

| Sample | Volume density (kg/m$^3$) | Young´s modulus (GPa) |
|---|---|---|
| SWCNT | 550 | 4.7 |
| rGO(th) | 690 | --- |
| SWCNT- GO | 820 | 6.7 |
| SWCNT- rGO | 270 | 0.3 |

### 3.3 Electronic transport

In Figure 5, the temperature dependences of the electrical conductivity $G(T)$ of pristine SWCNT (red line), and of pristine thermally-reduced graphite oxide - rGO(th) (mustard line) papers, and those of composites SWCNT-GO (purple line) and SWCNT-rGO (blue line) are presented in log-log scale. The log-log presentation enables to evaluate the electronic transport character through papers made of 1-dimensional SWCNT (b), 2-dimensional rGO(th) (d), and through the composites SWCNT-GO (c) and SWCNT-rGO (a). It is evident that, while curves a, b and c follow the same trend, curve d of the rGO(th) paper differs from the previous ones.

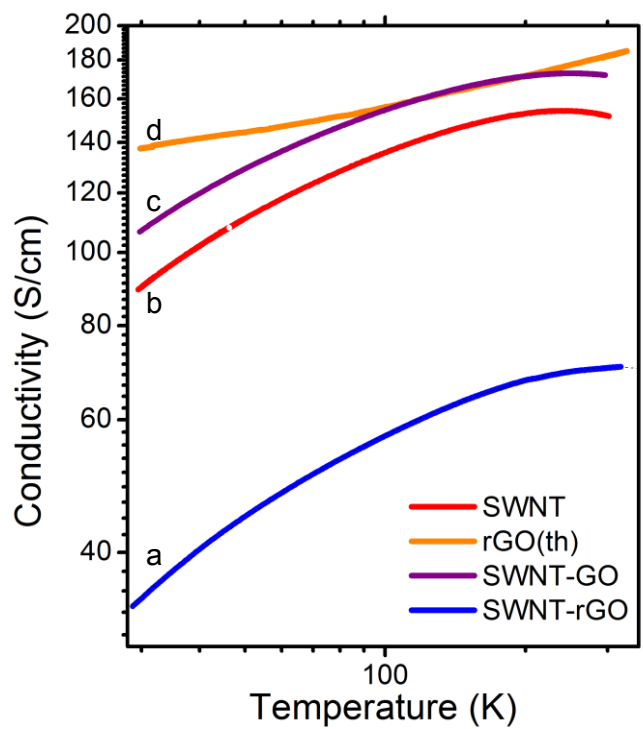

**Figure 5:** Temperature dependences of the electrical conductivity of: composite SWCNT-rGO (blue line-a); a SWCNT paper (red line-b); a composite SWCNT-GO (purple line-c); and a thermally reduced GO paper (rGO) (mustard line-d) – in log-log scale.

We observe that conductivity in SWCNT-GO is higher than that measured in the SWCNT paper. On the contrary, SWCNT-rGO shows a significant reduction in conductivity compared to that in SWCNT networks. Paper of rGO(th), besides the different character of $G(T)$, exhibits values of conductivity which are in the whole measured $T$-range higher than that for pristine SWCNT paper,

## 4 Discussion

### 4.1 Raman spectroscopy

A direct evidence of changes in the electronic structure of SWCNTs is provided by Raman spectroscopy [14,15]. Since SWCNT paper contains a mixture of tubes with various chirality and diameters that determine their electronic properties (semiconducting or metallic), varying excitation energies would visualize only those tubes where optical transition is in resonance with the laser excitation energy. In order to distinguish effects caused by charge transfer on semiconducting versus metallic tubes we use two excitation energies - 1.93 eV (633 nm) and 2.41 eV (514 nm).

The frequencies of the RBMs $\omega_{RBM}$ are directly related to the tube diameters. In order to identify those tubes which are in resonance with the two excitation energies, we compared the $\omega_{RBM}$ with the theoretical values calculated by Popov et al. [16]. The structural parameters and electronic character of the identified SWCNTs are summarized in Table II. Both prominent SWCNTs have similar diameters about 1.2 nm, typical for SWCNTs synthesized through the Laser Ablation (LA) method. However, their electronic character is different: at 514 nm, the observed Raman features represent semiconducting tubes, whereas at 633 nm, metallic tubes are in resonance.

**Table II:** Structural parameters of the prominent tubes identified from $\omega_{RBM}$ at two excitation energies.

| Wavelength (nm) | RBM $_{Exper.}$ (cm$^{-1}$) | RBM $_{Theor.}$ (cm$^{-1}$) | m | n | D (nm) | Character |
|---|---|---|---|---|---|---|
| 633 (1.93eV) | 195.8 | 195.9 | 13 | 4 | 1.21 | *metallic* |
| 514 (2.41eV) | 191.2 | 191.71 | 11 | 7 | 1.23 | *semiconducting* |

The RBM in both n- or p-doped nanotubes always exhibit a blue-shift [17]. When the semiconducting tubes are in resonance (Figure 4(a)), the RBM of SWCNT and SWCNT-rGO are almost identical, and the RBM of SWCNT-GO shows only a very small blue-shift in $\omega_{RBM}$. In Figure 4(d) displaying the RBM of metallic SWCNTs, the $\omega_{RBM}$ of SWCNT-GO (red line) is blue-shifted relative to the $\omega_{RBM}$ of the pristine SWCNTs (black line). This shift is a consequence of electronic doping. On the contrary, the $\omega_{RBM}$ of SWCNT-rGO (green line) has a small shift towards lower wavenumbers; this is a sign of "undoping", which is possible when the reference sample has been partially doped already in its pristine state [4].

The in-plane C-C stretching mode (G-mode) is displayed in Figure 4(b), (e). There is a significant difference in the features related to semiconducting versus metallic tubes. In Figure 4(b), the G-mode of semiconducting tubes has well distinguished two peaks originated from the C-C stretching (LO-mode) along the tube axis ($G^+$) and perpendicular (TO-mode) to it ($G^-$). Besides slightly increased intensity, there is almost no shift in $\omega_G$ of SWCNT-GO (red line) and of SWCNT-rGO (green line) relative to $\omega_G$ of SWCNTs (black line). On the other hand, the G-mode of the metallic tubes is much more complex (Figure 4(e)); its important feature is the wide asymmetric shape with a shoulder on the left side originating from electron-phonon coupling [18-22]. Comparing the spectra in Fig. 4e, it is evident that the spectral density is significantly redistributed for SWNT-GO compared to the other two spectra. The lowest energy component in SWNT-GO is blue-shifted by about 10 cm$^{-1}$ relative to that positioned at about 1540 cm$^{-1}$ in metallic pristine SWCNTs. Moreover, its line width and intensity decrease [21]. This G-mode development gives a clear evidence of electronic doping of metallic SWCNTs [21,22] in the presence of GO.

On the other hand, the G-mode of metallic tubes within SWCNT-rGO (Figure 5(e), green line) retains the features of pristine SWCNT, indicating that no electronic doping of nanotubes takes place.

## 4.2 Mechanical properties

If our interpretation from the previous paragraph of the Raman shift observed in SWCNT-GO by electronic doping were correct, not only Raman shift but electrical conductivity and elastic Young modulus would be affected [23] by the enhanced interactions induced by charge transfer between GO and metallic SWCNTs in their composite, or lack of the interactions in the SWCNT-rGO composite. In other words, the charge transfer would enhance also the mechanical properties of the paper and its volume density, and the opposite should happen when the interactions between the particles are weakened. The values in Table I indicate that in SWCNT-GO the enhanced conduction is accompanied by the highest value of Young´s modulus and by the densest structure, whereas in SWCNT-rGO, the suppressed conduction is accompanied by the lowest Young´s modulus and by the lowest volume density. Altogether, measurements of mechanical properties strongly support the interpretation based on electrostatic interaction through charge transfer between GO and SWCNTs.

## 4.3 Electronic transport

The electrical properties of SWCNT papers are well understood in the frame of Fluctuation-Assisted Tunneling (FAT) (Eq. 1):

$$R(T) = A\exp\left(-\frac{T_m}{T}\right) + B\exp\left(\frac{T_b}{T_s + T}\right) \qquad (1)$$

The second term represents fluctuation-assisted tunneling through barriers $k_B T_b$ [24]. In the region $T < T_s$, conductivity gradually becomes $T$-independent. This term gives the metallic extent of conductivity at the low-$T$ limit ($1/(B.\exp(-T_b/T_s))$) but, contrary to the "classical" metals, the tunneling mechanism, reflected in the second term of Eq. 1, causes $T$-dependence $R(T)$ to decrease with rising $T$. The first term takes into account 1-dimensional phonon back-scattering [25], where $k_B T_m$ is the energy of the back-scattering phonons. This term starts dominating at higher *T*, causing a change in the sign of derivative of the resistivity d*R*(*T*)/d*T*, from negative to positive. In other words, Eq. 1 describes a trade-off between the *T*-enhanced charge tunneling through energy barriers interrupting metallic areas in a network, against phonon scattering of the charge within the metallic areas, thus reducing conductivity as *T* rises. Yao et al. [26] show that the dominant scattering mechanism in individual SWCNTs at high electric field is emission either of optical phonon (1580 cm$^{-1}$) or of zone-boundary phonon (~1230 cm$^{-1}$). Also, in thick SWNT networks, T-dependences of conductivity fitted by the FAT model indicate similar scattering phonon energy corresponding to $T_m$ = 1830 K [2, 27, 28].

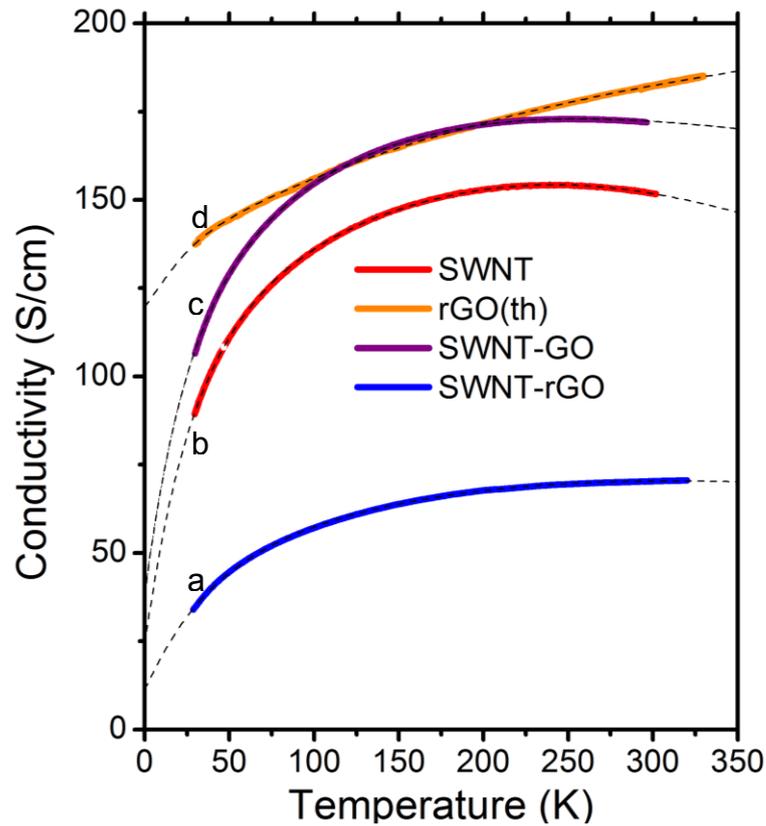

**Figure 6:** Temperature dependences of electrical conductivity of a composite SWCNT-rGO (blue line-a), of a SWCNT paper (red line-b), of a composite SWCNT-GO (purple line-c) and of a thermally reduced GO paper (rGO) (mustard line-d), in a linear scale. Experimental data are fitted with Eq. 1 (black dashed lines).

We analyze our data of transport measurements within the frame of the FAT model too. In order to visualize the fitting curves down to *T*=0, the experimental data are replotted in linear scale in Figure 6. The fitting parameters are collected in Table III.

**Table III** Fitting parameters of Eq. 1 for *G*(*T*) measured in various samples

| Samples | 1/A (S/cm) | $T_m$ (K) | 1/B (S/cm) | $T_b$ (K) | $T_s$ (K) |
|---|---|---|---|---|---|
| SWCNT | 35.39 | 1295.3 (905.8cm$^{-1}$) | 177.62 | 30.7 (2.7 meV) | 15.5 |
| rGO(th) | 477.7 | 94.9 (66.1cm$^{-1}$) | 445.9 | 299.5 (25.8 meV) | 227.6 |
| SWCNT-GO | 197.52 | 886.3 (616.4cm$^{-1}$) | 197.63 | 28.8 (2.5 meV) | 17.5 |
| SWCNT-rGO | 44.88 | 1219.4 (852.5cm$^{-1}$) | 83.40 | 46.0 (4.0 meV) | 23.0 |

In all cases the fitting agreement of the FAT model is excellent: The dashed fitting curves perfectly overlap with the experimental data in the measured *T*-range. Beyond this *T*-range, down to the low-*T* limit (T=0K), the FAT model predicts a metallic transport, retaining a final conductivity value given by 1/(**B**.exp($T_b/T_s$)).

In order to understand the mechanism of transport in the composite networks, let us first discuss transport in the reference samples. The major difference in the shape of the $G(T)$ curves of SWCNTs (b) and of rGO (d) in Fig. 6 comes from the entire properties of the 1-dimensional versus 2-dimensional nature of particles, and their interactions within networks.

The crystal structure of SWCNTs produced at very high temperatures by the LA method is well ordered; such a SWCNT behaves as a genuine quantum wire with ballistic transport of a current density of more than $10^9$ A/cm$^2$ [29]. On the other hand, the structural disorder present in a rGO sheet significantly reduces the intrinsic conductivity [9,10]. The disordered regions separating metallic regions represent here tunnelling barriers for charge transport.

When networks are formed, the contact area between SWCNTs available for charge tunnelling, from simple geometrical arguments, is much smaller than that between the planar rGO flakes.

In the case of SWCNTs, therefore, contact resistance between nanotubes represents the major factor limiting charge transport. As a consequence, at low $T$, $G(T)$ is small but steeply increases and then saturates at higher $T$, when thermal energy is sufficient to overcome most of the tunnelling barriers. At this point, intrinsic conductance of the entire nanotubes becomes dominating; since at higher $T$, scattering of electrons by emitting phonons affects the charge transport so that $G(T)$ decreases. The value of the fitting parameter $\mathbf{T_m}$ = 1295 K (905.8 cm$^{-1}$) for SWCNTs (Table III) corresponds to the energy of phonons at the peak in phonon density of states of SWCNT [31] that interact with conducting electrons at higher $T$.

The networks of rGO, to some extent, can be seen as disordered turbostratic graphite. As disorder strongly affects electronic transport, while for monocrystal graphite conductivity reaches the order $10^4$ S/cm [30], in our rGO(th) it is $10^2$ S/cm. Compared to the SWCNT networks, because of the large contact area between the rGO sheets, contact resistance is effectively reduced. On the other hand, as mentioned above, intrinsic resistivity of rGO is significantly higher due to the structural disorder. As **T$_m$** = 95 K corresponding to 66 cm$^{-1}$ (Table III), low energy acoustic phonons, apparently, cause electron scattering from relatively low *T*. The significantly lower value of **T$_m$** against that found in SWCNTs can be explained by the electron mean free path limited by the disordered islands in the graphene structure of rGO; it has been shown [ 32,9] that the typical distance between the disordered regions (and so the electron mean free path) is a few nanometers. Therefore, at this short distance electrons accelerating by electric field do not gather energy sufficient to emit high energy phonons. As a result of compromise between acoustic-phonon scattering and fluctuation assisted tunnelling through a large number of barriers, both inside the rGO sheets and between the sheets, *G*(*T*) moderately increases along the measured *T*-range. The significance of barrier size is reflected in the value of the fitting parameter **T$_b$** (Table III), which is of an order of magnitude larger than that for SWCNT, while (1/**A**+1/**B**) – the conductivity at .high *T*-limit – is four-times higher in rGO(th) than in the SWCNT paper.

Summing up the main factors controlling electronic transport in our reference samples:

- In SWCNT networks: high intrinsic conductivity of SWCNT versus poor coupling between nanotubes;

- In rGO networks: good coupling between the sheets versus poor intrinsic conductivity within the sheet.

Now we turn our attention to the composites: The intriguing similarity in the character of *G*(*T*) of both SWCNT-GO and SWCNT-rGO with *G*(*T*) of the pristine SWCNT paper (i.e. different from that of rGO(th)) as evident from Fig. 5, where *G*(*T*) is presented in log-log scale; this suggests that the carbon nanotubes are those who provide the charge transport in both composite samples. In fact, for SWCNT-GO it is expected since the conductivity of SWCNTs is many orders of magnitude higher than that of insulating GO. Based on the same argument we could also expect that 50 wt% of electrically insulating GO in the composite should significantly reduce conduction by separating carbon nanotube bundles with insulating barriers. Nevertheless, the opposite is true: We observe that the conductivity in the composite SWCNT-GO is even higher than that in the pristine SWCNT paper. This contradiction indicates that GO plays an active role in the charge transport through the SWCNT network. Likely, the covalently attached oxide groups to the graphene planes compensate for their unevenly distributed charge through transfer of charge from the metallic SWCNTs. In other words, GO sheets induce electronic doping of the metallic SWCNTs. The value of $T_m$ = 886 K (619.4 cm$^{-1}$) for SWCNT-GO composite indicates that a lower energy branches in phonon spectrum [31] cause charge scattering.

A similar effect of GO on electronic transport in a composite with conductive polymers was reported by Basavaraja et al [33]. The authors observe that the DC conductivity has higher value in the composite with GO than that in the polymeric matrix without GO. However, no interpretation of this observation is presented in the Ref. 33.

In the case of rGO, after reduction of the oxide groups, rGO becomes conductive. Therefore, the composite of rGO with SWCNTs should combine the transport behaviour of the pristine SWCNTs and rGO(th). It seems, however, that rGO here does not contribute to transport much. Since the conductivity of pristine rGO(th) paper reaches values higher than those of the SWCNT paper, it is not easy to figure out a reason why it fails, when rGO is mixed with SWCNTs. The resistivity in a single sheet of rGO is typically three orders of magnitude higher ($10^6$ Ω) [10] than that in a single sheet of exfoliated graphene ($10^3$ Ω) [5, 6]. There is no charge transfer between rGO and SWCNTs that could induce strong electrostatic interactions. Mixing SWCNTs with rGO presumably even reduces charge tunnelling between the SWCNT bundles which become separated by much less conductive rGO flakes, and in addition randomly mixed SWCNTs reduce the contact area between the rGO sheets. The scattering phonons even though very similar to those in pristine SWCNTs are slightly softened: for SWCNT-rGO composite $T_m$ = 1219 K (852.5 cm$^{-1}$). A similar trend of mode softening in SWCNT-rGO is also seen in the red-shift of RBM and 2D modes of Raman spectra (Fig. 4d,f)

Our argument in order to explain the conductivity behaviour is based on the charge transfer between GO and SWCNT and its absence between rGO and SWCNT. The SEM images in Fig. 1c,d show the morphology of the densely packed SWCNT-GO against the loosely packed SWCNT-rGO. Strong evidence of different interactions within two composites, in addition, is provided by results of Young modulus measurements and Raman spectra discussed above.

## 5. Summary

We show that the presence of electrically insulating GO within a SWCNT network strongly enhances electrical conductivity, whereas reduced rGO, even though electrically conductive, suppresses electrical conductivity within a composite network with SWCNTs. In both cases of composite networks, the transport mechanism clearly retains the same features as that of the pristine SWCNT network, whereas it differs from the highly metallic transport behaviour of the pristine rGO(th) network.

In addition, measurements of Young´s modulus and of Raman spectra strongly support our interpretation of the "indirect" role of the oxide groups, present in GO within the SWCNT-GO composite, through electronic doping of metallic SWCNTs. The charge transfer between GO and SWCNTs is proven by the blue-shift of RBM- and G-modes in metallic tubes and it is reflected in increased volume density, electrical conductivity and in increased elastic modulus of the SWCNT-GO composite, compared to those in the pristine SWCNT network. The same experiments prove absence of the charge transfer between rGO and SWCNTs showing the red-shift of some Raman modes in metallic tubes and decrease in volume density, electrical conductivity and in elastic modulus of the SWCNT-rGO composite.

**Acknowledgements**

This work was supported by funding from the European Union`s Seventh Framework Programme (FP7/2007-2013) under grant agreement n°266391 (Electrograph). M.H. acknowledges support from the project VEGA 1254/12. J.C.M. acknowledges support from the Austrian Science Fund (FWF): I1283- N20 . We thank Pavol Kováč, Peter Martis and Jozef Novák for their assistance at electrical measurements.